\documentclass[]{spie}  

\usepackage{amsmath,amsfonts,amssymb}
\usepackage{graphicx}
\usepackage{booktabs}
\usepackage{subcaption}
\usepackage{paralist}
\usepackage{enumitem}
\usepackage{array}
\usepackage{geometry}
\usepackage{ulem}
\usepackage{xcolor}
\usepackage{siunitx}
\DeclareSIUnit{\belmilliwatt}{Bm}
\DeclareSIUnit{\dBm}{\deci\belmilliwatt}

\usepackage[colorlinks=true, allcolors=blue]{hyperref}

\title{The CRS: a scalable full-stack control system for Microwave Kinetic Inductance Detectors}

\author[a,b]{J. Montgomery}
\author[a,b]{W. Avelino}
\author[a]{M. Dobbs}
\author[a]{J. Letang}
\author[b]{M. Rouble}
\author[a]{S. Savchyn}
\author[a]{G. Smecher}

\affil[a]{t0.technology Inc., Montreal, Canada}
\affil[b]{McGill University, Montreal, Canada}

\authorinfo{Further author information, send correspondence to Dr. Joshua Montgomery\\E-mail: Joshua@t0.technology}

\begin{document} 
\maketitle

\begin{abstract}
The t0.technology Control and Readout System (CRS) is a modular microwave control and readout system for mm-wave and radio astronomy, THz imaging, noise radar, and superconducting qubit control. The configuration discussed in this work implements firmware for readout of microwave Kinetic Inductance Detector (KID) arrays. The CRS can operate 4,096 KIDs over \SI{2.5}{\giga\hertz} of complex bandwidth between 0--\SI{10}{\giga\hertz}, typically allocated across four independent RF chains at 1,024x multiplexing and \SI{625}{\mega\hertz} of complex bandwidth each. Every CRS can operate as a standalone unit or collectively within one or more backplane-enabled subracks that distribute power, clocking, and synchronization, scaling to an arbitrary number of channels. Each fully populated subrack supports arrays of more than 65,000 KIDs. The signal processing and control software supports recent innovations in multi-probe measurements and dynamic feedback modes, which are described in Rouble et al. (2024, these proceedings). The CRS has recently been selected as the new baseline readout system for the proposed South Pole Telescope instrument, SPT-3G+.\cite{Anderson2022} We present the hardware design, firmware capabilities, open-source control and data acquisition software, and the first laboratory characterization measurements.

\end{abstract}


\section{INTRODUCTION}
\label{sec:intro}  

\begingroup

The t0.technology Control and Readout system (CRS) is a multi-functional platform for low noise scientific systems that operate across 0--\SI{10}{\giga\hertz} and require significant signal processing. It is specifically designed to address the life-cycle requirements to develop and field large-scale instrumentation for astrophysics and particle-physics. This document highlights the system in the context of operating Kinetic Inductance Detectors (KIDs).

KIDs are band-pass resonators made from inductive thin-film superconductors.\cite{Day2003}
Their electrical properties are a strong function of the density of bound electron quasi-particles within the film, and are sensitive to small depositions of energy from light (photons) or heat (phonons) that temporarily break those bound pairs back into the constituent electrons. When the quasi-particle density in the film fluctuates, the complex impedance of the resonator changes in a measurable way, making them effective detectors across the electromagnetic spectrum, particularly for low energy millimetre-wave imaging and phonon-mediated particle physics. When used for imaging, KIDs operate at the photon-noise limit, such that the dominant contribution to the system noise is the Poisson statistics of incoming photons. The development of ever more powerful instruments using these devices is also the development of operating ever more densely populated arrays of them.
KIDs can be multiplexed by manufacturing them to have different resonant frequencies and placing them in parallel on a single microwave feed-line, where complex impedance for each resonator is readout through S21 measurements. The control and readout system is heavily implicated in this kind of design, and the scale at which these devices can be multiplexed. Foremost, it:
\begin{compactenum}
    \item Must not be the limiting factor dictating achievable multiplexing density or bandwidth.
    \item Must not spoil photon-noise limited performance.
\end{compactenum}
KIDs measurements are often used to reconstruct signals that are statistical in nature, or subject to large systematic fluctuations in foreground signals. The control and readout system:
\begin{compactenum}
    \setcounter{enumi}{2}
    \item Must maintain a linear response, with strictly controlled cross-talk, and a wide dynamic range.
    \item Must further linearize cryogenic signal path elements through digital feedback.
\end{compactenum}
From a life-cycle perspective, KIDs design and operation is rapidly evolving, and this research and development phase requires control and readout systems that:
\begin{compactenum}
    \setcounter{enumi}{4}
    \item Provide detailed visibility into the superconducting system, particularly measurements of dynamic behavior.
    \item Have an open API that can be customized by researchers, who are often using novel device configurations.
\end{compactenum}
Following the development phase, a deployed instrument requires control and readout systems with:
\begin{compactenum}
    \setcounter{enumi}{6}
    \item High reliability and system up-time in harsh environments.
    \item Effective integration with other infrastructure, such as Masers, GPS timing, or high-performance computing and data acquisition.
\end{compactenum}

The CRS was designed for the full life-cycle requirements of instruments that use KIDs. The sections that follow introduce the hardware, signal processing firmware, and software, and conclude with examples of integration with third-party control software and KIDs characterization measurements.
\begin{figure}[h!]
    \centering
    \begin{subfigure}[t]{0.45\textwidth}
        \centering
        \includegraphics[width=\textwidth]{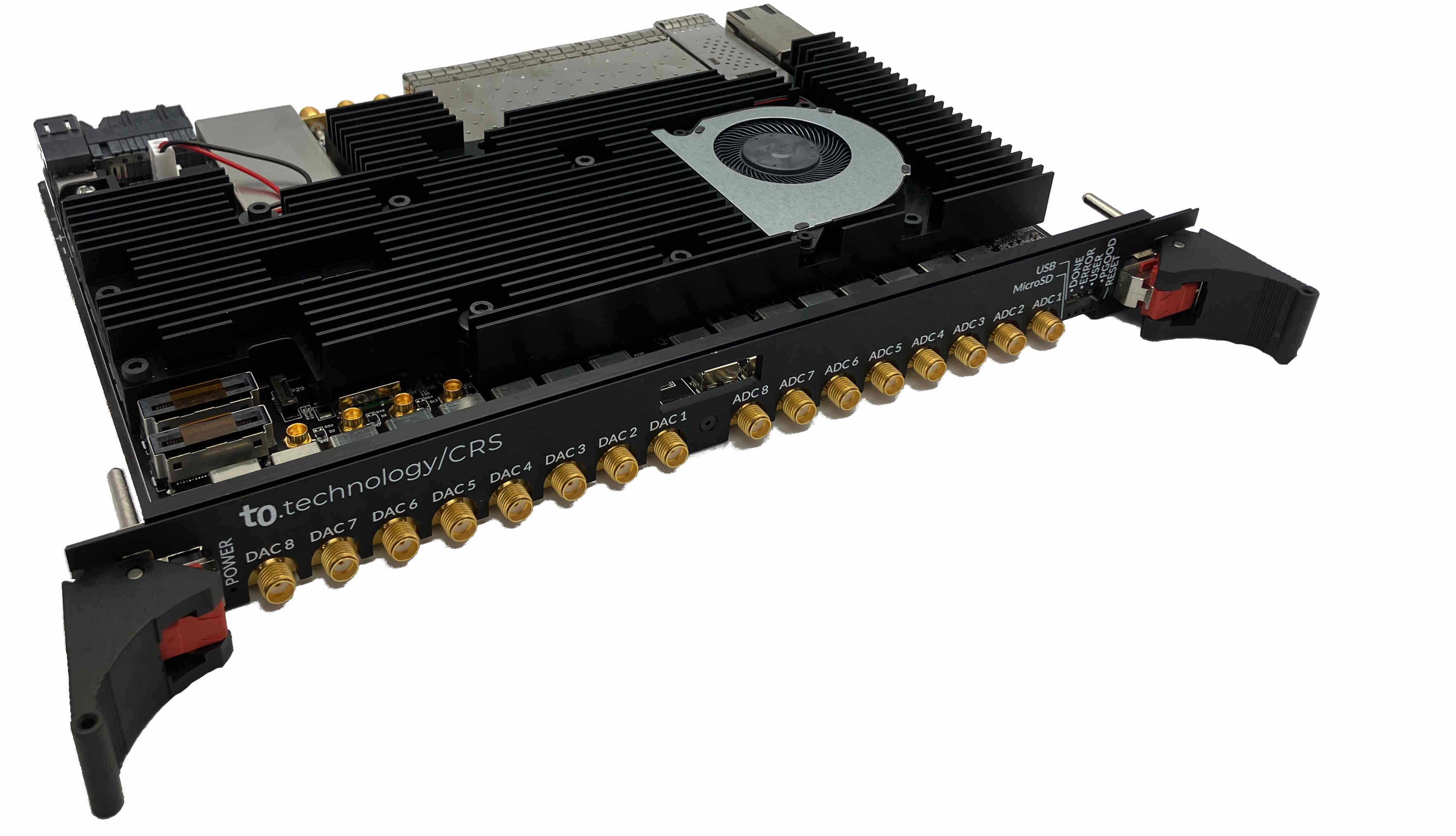}
        \caption{A self-contained CRS board, completely assembled and ready for insertion into either a single-board enclosure or a backplane.}
    \end{subfigure}
    \hfill
    \begin{subfigure}[t]{0.45\textwidth}
        \centering
        \includegraphics[width=\textwidth]{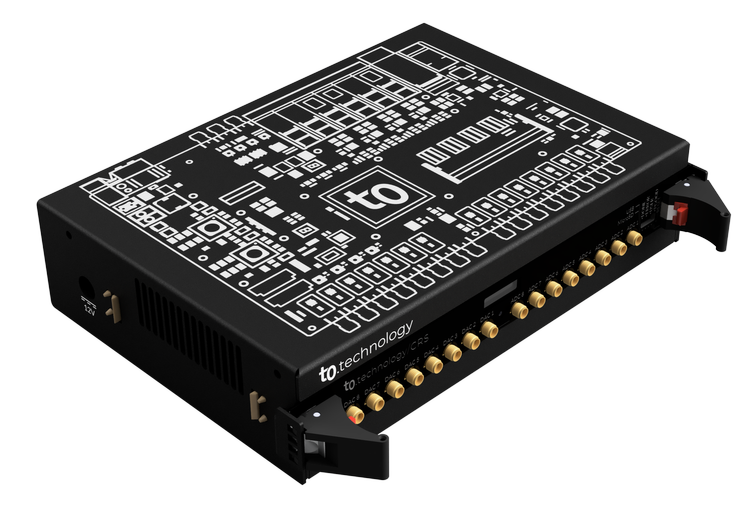}
        \caption{A single board enclosure for laboratory-scale operations.}
    \end{subfigure}
    \vfill
    \begin{subfigure}[t]{0.45\textwidth}
        \centering
        \includegraphics[width=\textwidth]{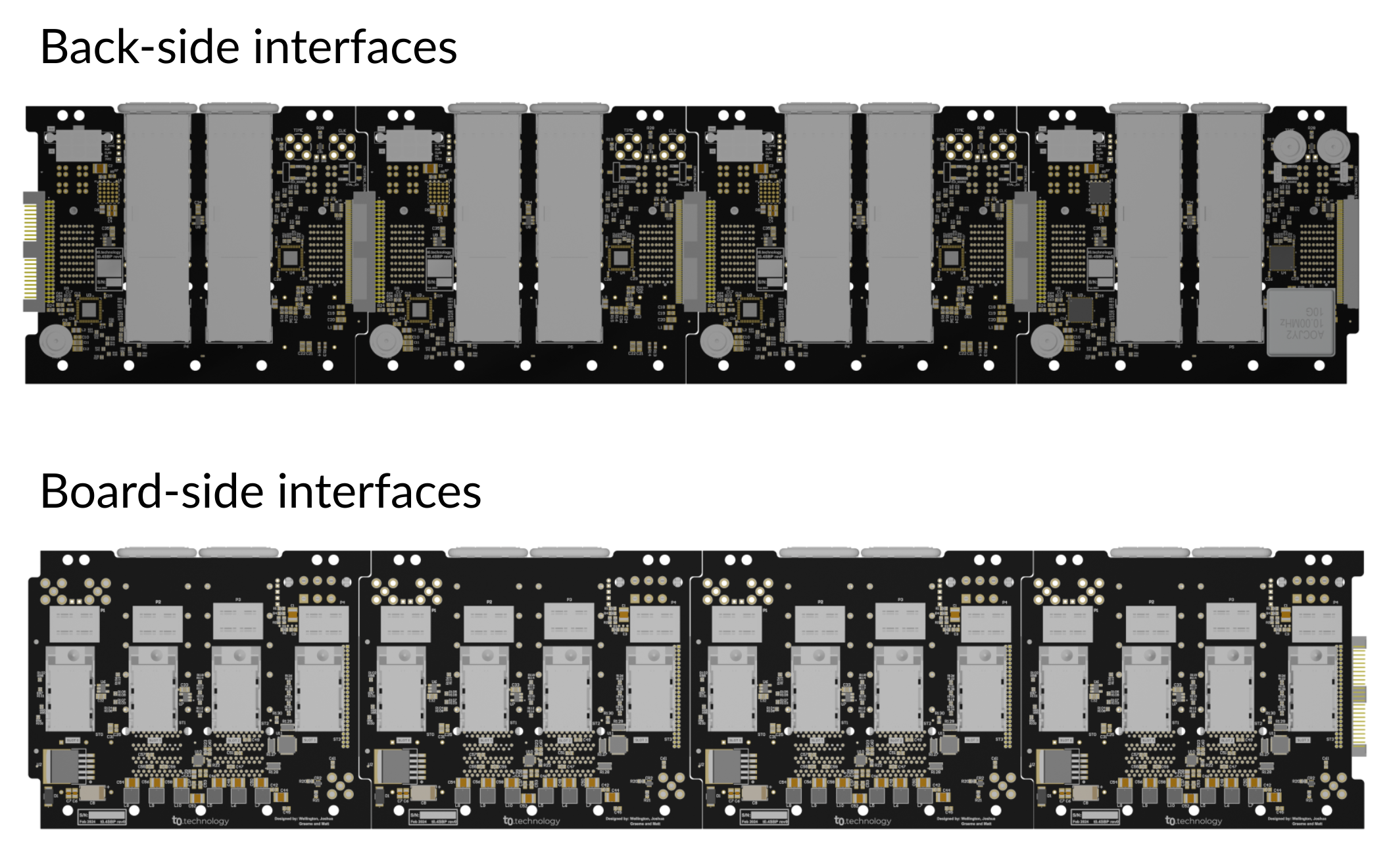}
        \caption{The backplane assembly distributes power, clocking, timing, and high-speed networking between 16 CRS boards.}
    \end{subfigure}
    \hfill
    \begin{subfigure}[t]{0.45\textwidth}
        \centering
        \includegraphics[width=\textwidth]{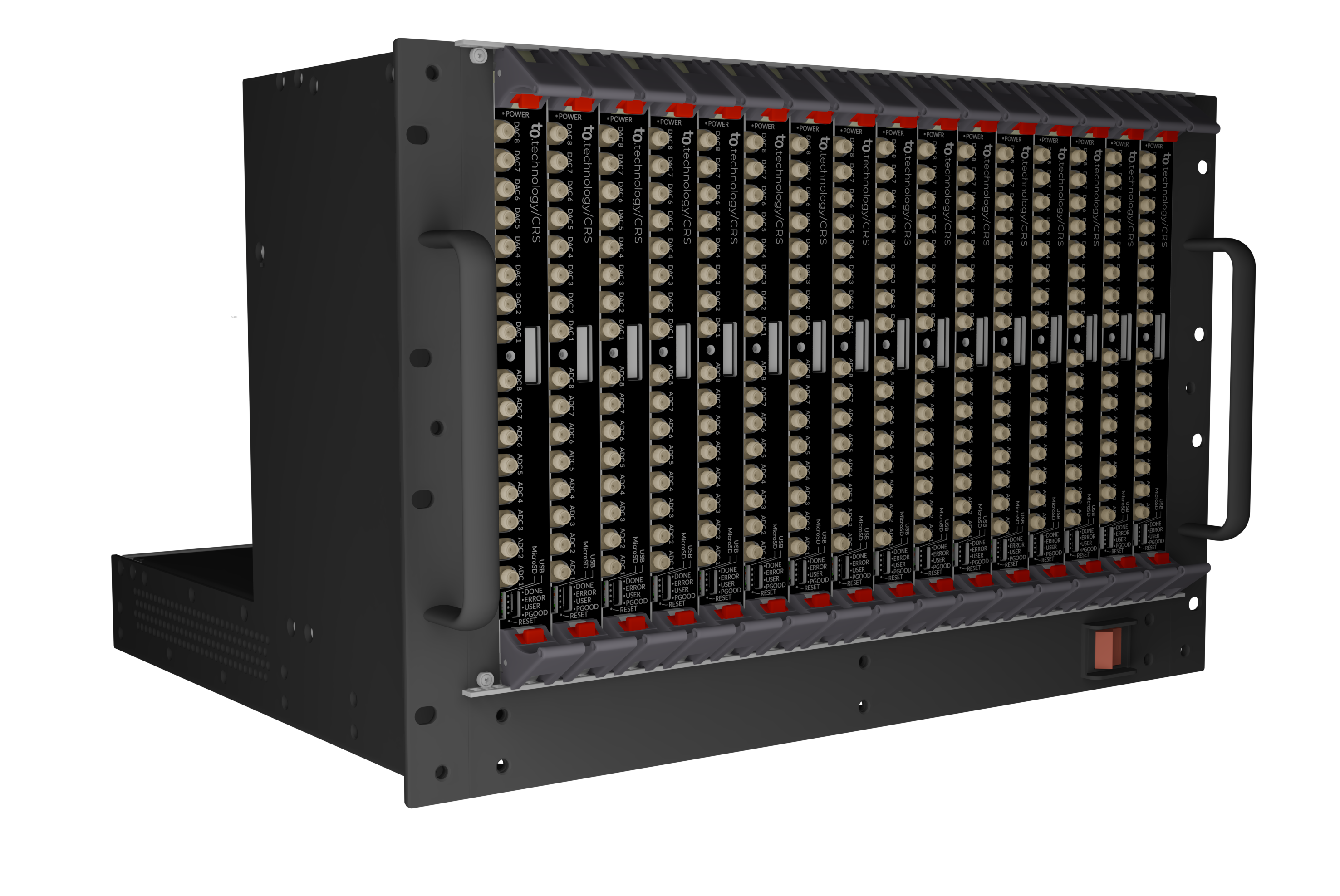}
        \caption{A fully assembled subrack, which integrates a power supply and fan tray, the backplane assembly, and up to 16 CRS boards. A 4-slot backplane is also available.}
    \end{subfigure}
    \vspace{1em} 
    \caption{Overview of the CRS configurations and assemblies.}
    \label{fig:combined}
\end{figure}

\section{THE CRS MOTHERBOARD HARDWARE}

Each CRS board is built around the 3rd generation AMD Radio Frequency System on a Chip (RFSoC), which incorporates a Field Programmable Gate Array (FPGA); ARM CPUs; a bank of analog RF converters; and high-speed digital data transceivers. The RFSoC is hosted on a 6U \SI{160}{\milli\meter} PCB that breaks out the peripheral interfaces. The digital design of the CRS includes a requirement that all oscillators are phase-synchronized to a single \SI{10}{\mega\hertz} master clock, ensuring no free-running clocks across any peripheral. This is a departure from traditional digital signal design of general purpose RFSoC systems, and is necessary to eliminate 1/f noise contributions due to pick-up from uncorrelated clock drifts. The analog board design is segregated from high-speed digital signals, and preserves the native analog bandwidth of the RFSoC with a wide-band \SI{50}{\ohm} balun to AC-couple the RF inputs and outputs. The relevant interfaces are shown in Figure \ref{fig:crs-xray}, and summarized in the sections below.
\begin{figure}[h]
    \centering
    \includegraphics[width=0.8\linewidth]{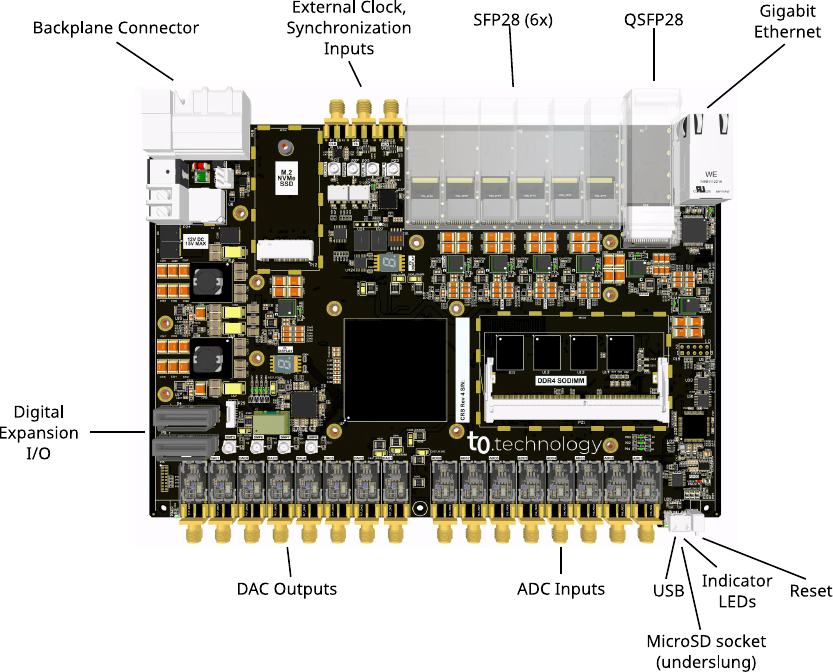}
    \vspace{1em}
    \caption{A CRS board with heat sink and front panel removed and annotations for major interfaces and connectors.}
    \label{fig:crs-xray}
\end{figure}

\subsection{Analog specifications}

Each CRS board hosts 8 separate RF-ADC and RF-DAC channels. In the present firmware architecture a subset of these are active simultaneously, typically 4 independent RF chains, depending on the target bandwidth (see Section \ref{sec:firmware}). In all cases the active ports can be dynamically selected, such that a CRS can be used to test up to 8 separate KIDs-coupled transmission lines without manually switching cabling.

The analog performance is well described by the RFSoC device itself\footnote{Detailed specifications: \url{https://docs.amd.com/r/en-US/ds926-zynq-ultrascale-plus-rfsoc}}, and end-to-end measurements have returned dynamic range, noise, and bandwidth specifications consistent with the performance of the chip, after accounting for approximately \SI{2}{\decibel} of insertion loss at the balun. A summary of the analog bandwidth and dynamic range is given in Table \ref{tab:adc_dac_specs}.
\begin{table}[h]
    \centering
    \begin{tabular}{lll}
        \textbf{Parameter} & \textbf{Specification} & \textbf{Notes} \\
        \toprule
        \multicolumn{3}{l}{\textbf{ADC}} \\
        Bits & 14-bit & \\
        Sampling Frequency & 5 GSPS &  \\
        Analog bandwidth & 6 GHz (3dB) & \\        
        Digital Step Attenuation (DSA) & Up to 20dB & Adjustable in 1dB increments \\
        Full-scale range & 1 dBm (DSA=0 dB) & Up to a 14.6 dBm maximum \\
        \midrule
        \multicolumn{3}{l}{\textbf{DAC}} \\
        Bits & 14-bit &  \\
        Sampling Frequency & 5 GSPS (current firmware) & 9.85 GSPS (hardware capable) \\
        Analog bandwidth & 6 GHz (3dB) & \\        
        Full scale output & –18.5 dBm to 6.5 dBm & Dynamically variable \\
        \bottomrule
    \end{tabular}
    \vspace{1em} 
    \caption{ADC and DAC parameters.}
    \label{tab:adc_dac_specs}
\end{table}

\subsubsection{Noise performance}

The data-sheet specifications of the RFSoC chip do not include performance metrics in all of the relevant regimes for operating KIDs. Notably, the data-sheet Noise Spectral Density of \SI{-145}{\decibel\per\hertz} is given for kHz-scale offsets, rather than the most relevant near-carrier performance for KID operation. The noise spectral density in the near-carrier regime between 0--\SI{100}{\hertz} is shown in Figure \ref{fig:loopback-psd}.

\begin{figure}[h!]
    \centering
    \includegraphics[width=0.65\linewidth]{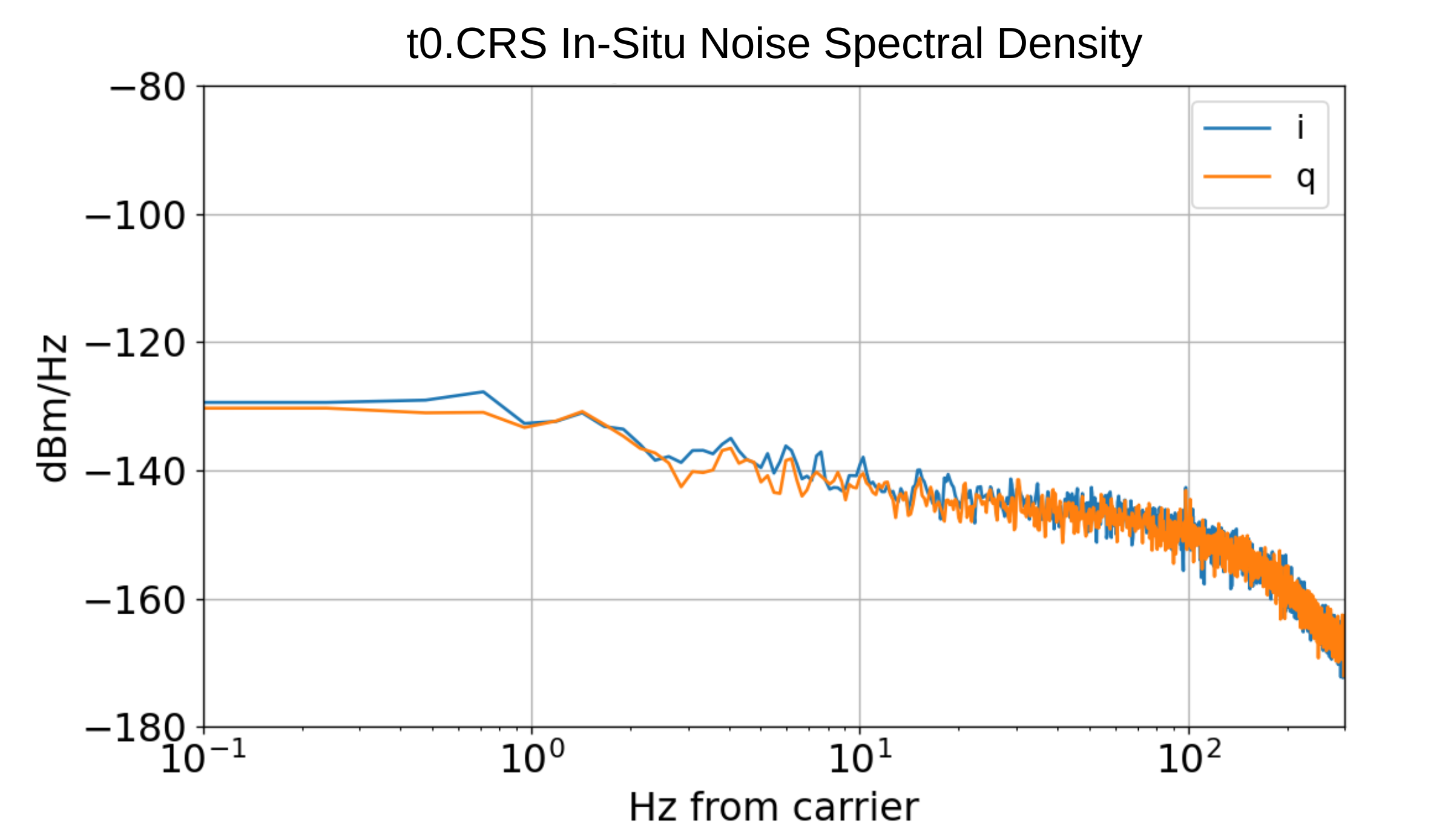}
    \vspace{1em}
    \caption{Loop-back measurement of the t0.CRS noise spectral density referred to the input of the ADC, using a tone power of \SI{-45}{\dBm} and frequency of $\sim$\SI{500}{\mega\hertz}. These data were taken at a sampling frequency of 596 samples per second. The roll-off at higher frequencies is due to a digital anti-aliasing filter.}
    \label{fig:loopback-psd}
\end{figure}

\subsubsection{Higher Nyquist regions}

To preserve the performance of the system at higher Nyquist regions, no Nyquist band-defining filters are incorporated onto the board. AMD provides several mix modes that allow for improved performance when generating higher Nyquist signals or digitizing from higher Nyquist zones. These have been examined in detail for similar systems using the same RFSoC, and operated as high as \SI{10}{\giga\hertz}\cite{Park2022}. The CRS microwave firmware exposes a selection between NZD and RC mix modes, and the CRS has operated KIDs up to \SI{5.7}{\giga\hertz}.

\subsection{Clocking and timestamping}

Each CRS board includes an onboard crystal oscillator for laboratory and standalone use, and also accepts external \SI{10}{\mega\hertz} timing signals via the backplane or auxiliary SMA input. These are prioritized automatically, such that the auxiliary SMA is used if present, followed by the backplane input, and finally the onboard oscillator. The on-board oscillators are shut down when not used.
Data is time-stamped using an external Inter-Range Instrumentation Group timecode (IRIG)-B source from a GPS receiver received via the backplane or rear SMA port, and can also be time-stamped with a a mock IRIG generator built into the CRS.

\subsection{Digital data transfer}

Each CRS board has a single 1\,Gbps Ethernet interface and 16 independent high-speed 25\,Gbps GTY transceivers allocated across several interface protocols, summarized in Table \ref{tab:crs_interfaces}. Of these, 150\,Gbps are reserved for data transfer across a backplane, and 150\,Gbps are currently only used for radio astronomy applications; however, the 100\,Gbps QSFP28 interface is optionally available for high bandwidth streaming of KID data.

\begin{table}[h!]
    \centering
    \begin{tabular}{lll}
        \textbf{Interface} & \textbf{Data Rate} & \textbf{Description} \\
        \toprule
        Gigabit Ethernet RJ45 & 1 Gbps & Standard interface for all commanding and data transfer \\
        1x QSFP28 & 100\,Gbps & Optional interface for full data-rate streaming \\
        6x SFP28 & 150\,Gbps total & Used for radio astronomy applications \\
        Backplane GTY & 150\,Gbps total & Used for radio astronomy applications \\
        \bottomrule
    \end{tabular}
    \vspace{1em} 
    \caption{Networking and data transfer interfaces.}
    \label{tab:crs_interfaces}
\end{table}

\subsection{Peripheral devices and I/O}

The CRS includes several peripheral I/O interfaces that are user-configurable. Table \ref{tab:io_interfaces} includes a complete list of extended I/O.
\begin{table}[h!]
    \centering
    \begin{tabular}{ll}
        \textbf{Interface} & \textbf{Description} \\
        \toprule
        1x DC-coupled SMA & Configurable as an input or an output \\
        4x SMP & DC-coupled I/Os \\
        4x SMP & AC-coupled clock/synchronization outputs \\
        84x high-speed LVDS & Two vertical sliver SFF-TA-1002 connectors \\
        micro-USB & Used for JTAG over USB for low-level shell access \\
        M.2 slot & Non-volatile storage or PCIe expansion \\
        Soldered RAM & 8\,GB 64-bit DDR4 \\
        DDR SODIMM & DDR4 260-pin for adding up to 32\,GB of additional RAM\\
        \bottomrule        
    \end{tabular}
    \vspace{1em} 
    \caption{I/O interfaces and peripherals.}
    \label{tab:io_interfaces}
\end{table}

\subsection{Onboard processing subsystem}

The RFSoC includes two ARM CPUs (a \SI{1.2}{\giga\hertz} Quad-core Cortex-A53 and \SI{500}{\mega\hertz} Dual-core Cortex-R5F) with direct memory with the FPGA subsystem.
These CPUs run a complete embedded Linux distribution.\footnote{A Board Support Package (BSP), including kernel and userspace software, is available at \url{www.github.com/t0}.}
The Linux subsystem is the interface between an external control computer and the signal processing functions within the FPGA. This interface is made up of a C++ API with Python bindings. Each CRS hosts a local JupyterLab instance, which provides a convenient way to control the CRS interactively.
When connected to a network, the CRS launches the JupyterLab instance with a local domain name corresponding to the board serial number, or subrack/slot position.

Because the Linux subsystem has access to the peripheral I/O devices and runs a persistent file system, it can host custom routines written in high-level programming languages such as C/C++ or Python to expand integration of the CRS with external equipment, perform low-latency algorithmic operations, customize the information displayed by the front-panel OLED screen, or facilitate other data operations such as initial data analysis or buffered data storage.

\subsection{Power and thermal management}

When operated in single-board configuration, the CRS is powered by 12\,VDC through either binding posts or a 2.5x5.5\,mm barrel adaptor compatible with a bundled AC/DC converter.
Each CRS draws between \SI{50}{\watt} and \SI{80}{\watt} of power, depending on the configuration. This power is dissipated with a conformal heat sink containing an integrated fan. The heat sink and fan remain affixed to the board whether operated in a single-unit enclosure or subrack.

\section{THE BACKPLANE ASSEMBLY}

The backplane distributes power, clocking, and IRIG time stamps to each CRS board. 
It also implements a high speed digital data transfer mesh between the boards.
The backplane occupies 7U of rack space, with boards in the upper 6U and a power supply/fan tray in the lower 1U. The power supply (\SI{1.2}{\kilo\watt}) and fan tray (110\,CFM) are sufficient to power and cool all 16 CRS boards under ordinary operating conditions. A higher-capacity PSU is also available where more power or margin is required (e.g. for deployments at altitude that require significant de-rating).

The backplane uses a segmented architecture, driven by the requirement for 25\,Gbps data links that must traverse the entire length of the subrack.
Implementing the data links exclusively in PCB substrate would drive the system towards higher cost and complexity, requiring expensive low-loss materials and many layers.
Instead, the backplane is subdivided into 4 segments that each service 4 CRS boards across a Megtron-4 substrate, with substrate links between those four boards. Low-loss QSFP patch cables are then used for data links between the segments. The only signals distributed directly across all segments are clocking, timing, and synchronization. The assembled segments are shown in Figure \ref{fig:bp-interfaces}.
\begin{figure}[h]
    \centering
    \includegraphics[width=0.85\linewidth]{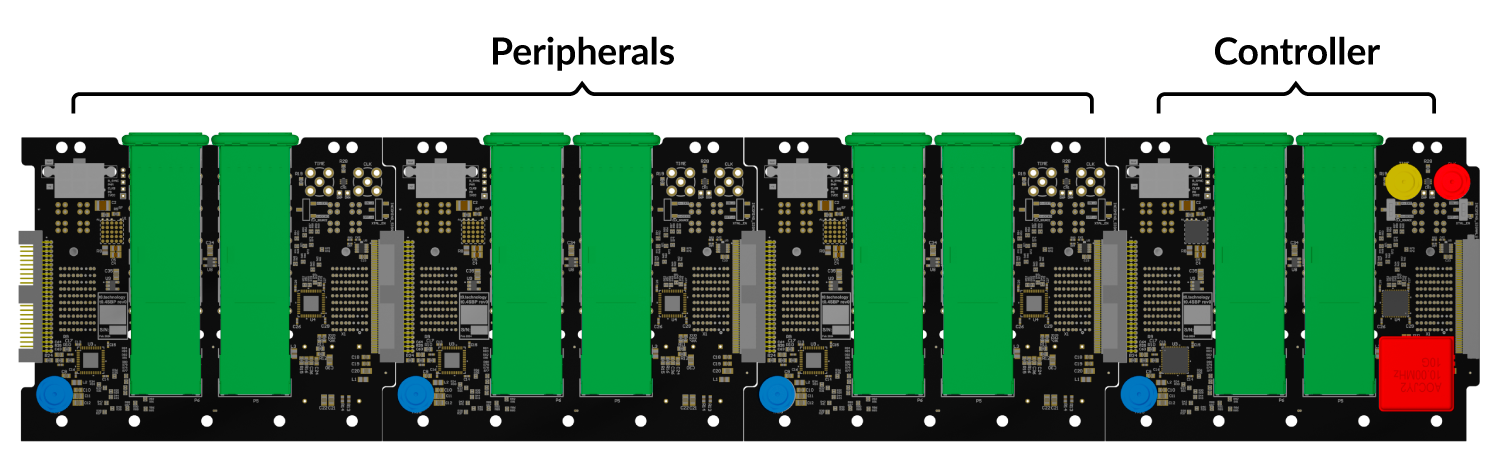}
    \vspace{1em} 
    \caption{The extended interfaces of the backplane assembly. Each assembly is built from 4 segments, consisting of a controller segment and three peripherals. The red indicates a \SI{10}{\mega\hertz} OCXO clock that is distributed across the backplane and bidirectional SMA port, used to either export the OCXO or import an alternative \SI{10}{\mega\hertz} clock. Yellow shows an IRIG SMA input for timing distribution. Green indicates double-height QSFP cages, three of which are active per segment for data transfer between the segments. Blue is a user-configurable bidirectional SMA port that is distributed within each segment and can be used for synchronization signals. 12\,VDC power is distributed from the subrack via Molex connectors on each segment.}
    \label{fig:bp-interfaces}
\end{figure}

\subsection{Clocking, timing, synchronization}

The backplane includes an Oven Controlled Crystal Oscillator (OCXO) with excellent phase noise performance.\footnote{Abracon AOCJY2-10.000MHZ-E.} The OCXO is distributed to all CRS boards and routed to an external SMA, allowing the CRS system to serve as the central clock to synchronize external equipment. Alternatively, the OCXO can be disabled and the SMA can be used as an input to distribute an external \SI{10}{\mega\hertz} clock, such as a high precision hydrogen maser clock commonly used in astronomical observatories. An IRIG SMA input also distributes timing across the full backplane. 
There is an un-allocated bidirectional SMA that is shared among the boards within each segment.


\subsection{Addressing and monitoring}

Each backplane includes temperature monitors and EEPROMs to allow for software identification of the backplane serial number and each CRS slot. Installed CRS boards declare their mapping through mDNS, and can be addressed uniquely by either board serial number or subrack serial number and position. This mapping ensures the control software remains agnostic to the specific hardware units used for a given topology, and can auto-discover the installed topology.

\subsection{Data shuffle}
The six 25\,Gbps GTY data transceivers per CRS board allocated to the backplane form a ``full mesh'' between all units in a subrack, as required for radio astronomy and FFT beam-forming. The further SFP28 and QSFP28 interfaces on each CRS board allow that mesh to grow to include up to 16 subracks (256 CRS boards). The mesh design, characterization, and corresponding signal path will be described in more detail in a future work, but has demonstrated lossless data transmission at 25\,Gbps across all links.

\section{THE rfmux MICROWAVE CONTROL FIRMWARE}
\label{sec:firmware}

Kinetic Inductance Detectors (KIDs) have different readout requirements depending on the science objective. 
Instruments such as imaging surveys or spectrometers measure continuous integrated power deposition, with target signals occupying kHz of bandwidth for each detector channel. These instruments often place channels within an octave of bandwidth to mitigate contamination from inter-modulation distortion products, requiring dense spacing and high multiplexing factors.
In contrast, particle physics, dark matter detection, or photon-number-resolving instruments measure discrete energy depositions by resolving individual pulses, and require detector bandwidths in the MHz range. These systems typically feature lower multiplexing density but sometimes occupy several octaves of bandwidth.\cite{Fruitwala2020}. Both of these configurations can be augmented with different real-time analog and digital feedback techniques to improve scalability and linearize detector response.

The \texttt{rfmux} microwave control firmware is designed to support all of the above requirements through different bandwidth, channel density, and data streaming modes. 
It specifically couples digitization and synthesis within the signal processing logic, allowing low-latency implementations of digital feedback schemes. 
The firmware architecture and operational configurations are described in Section \ref{sec:signal-path}; Section \ref{sec:streaming} provides additional detail regarding data streaming bandwidths; and Section \ref{sec:feedback} describes the basic structure of the digital feedback path.

\subsection{Signal path}
\label{sec:signal-path}

The fundamental building blocks of the firmware signal path are parameterized by:
\begin{description}[style=unboxed, leftmargin=0.5cm, labelsep=0.2cm, itemsep=0pt, parsep=0pt, topsep=0pt]
    \item[Module count:] the number of independent DAC/ADC interface pairs simultaneously operable;
    \item[Multiplexing (mux) factor:] the number of sinusoidal probe channels per I/O module;
    \item[Instantaneous bandwidth:] the bandwidth over which channels can be placed per I/O module;
    \item[Channel bandwidth:] the complex (I,Q) bandwidth continuously captured for each channel.
\end{description}
An I/O module operates one complete RF chain, which includes synthesis through an RF-DAC and digitization through an RF-ADC.
The standard configuration uses four active I/O modules, each configured for 
an instantaneous complex bandwidth of \SI{625}{\mega\hertz} over which 1,024 channels can be placed.
The wideband configuration consolidates the full instantaneous bandwidth and channel count within a single I/O module (see Table \ref{tab:kids_config} for configuration comparison).

\begin{table}[h!]
    \centering
    \begin{tabular}{lll}
        \textbf{Parameters} & \textbf{Default} & \textbf{Wideband} \\
        \toprule
        Module count & 4 & 1 \\
        Instantaneous bandwidth & \SI{625}{\mega\hertz} & \SI{2.5}{\giga\hertz} \\
        Mux factor & 1,024 & 4,096 \\
        \bottomrule
    \end{tabular}
\vspace{1em} 
\caption{Comparison of two configurations for the \texttt{rfmux} microwave control firmware.}
\label{tab:kids_config}
\end{table}

The signal processing architecture uses coupled Polyphase Filter Bank (PFB) synthesis and demodulation blocks, which can be allocated across the analog I/O interfaces. Each PFB processes 1,024 independent channels across \SI{625}{\mega\hertz} of complex bandwidth. 
Each channel can be independently parameterized by frequency, amplitude, and phase, and allocated arbitrarily within the instantaneous complex bandwidth with sub-mHz frequency resolution. The center of that bandwidth is set via programmable complex Numerically Controlled Oscillators (NCOs) and decimation/interpolation stages that interface with RF-DAC and RF-ADCs. The internal structure of the PFB blocks operate at 2.44\,MSPS, which sets the maximum streamed (I,Q) channel bandwidth and the maximum bandwidth for feedback between input and output signals.
A block diagram of the signal path in the standard configuration is shown in Figure \ref{fig:signal-path}.

\begin{figure}[hbtp]
    \centering
    \includegraphics[width=1\linewidth]{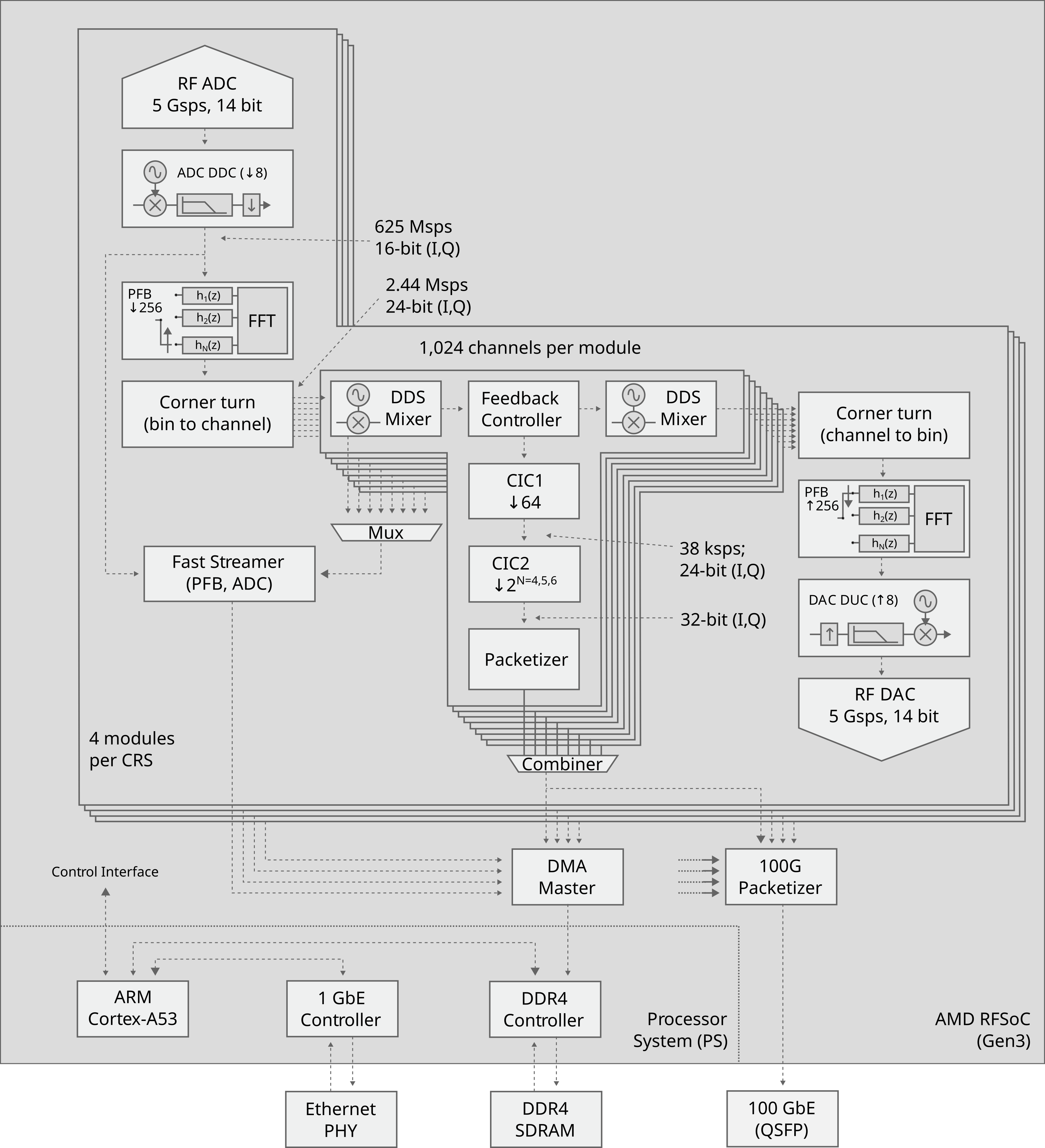}
    \vspace{1em} 
    \caption{Detailed block diagram for the signal path of the \texttt{rfmux} firmware in the default configuration.}
    \label{fig:signal-path}
\end{figure}

\subsection{Data streaming and capture}
\label{sec:streaming}
All channels are channelized and demodulated as complex (I,Q) time-ordered data at a native bandwidth of 2.44\,MSPS.
This native bandwidth is reduced through a series of CIC filters with a selectable decimation rate. The available down-sampling rates accommodate continuous streaming of the full 4,096 channels through a 1\,Gbps Ethernet interface via UDP multicast, with a maximum rate of 2.384\,kSPS. The 100\,Gbps QSFP28 interface can be used for faster sampling rates, either to stream the native 2.44\,MSPS (I,Q) data for up to 512 channels, or raw 625\,MSPS (I,Q) outputs of the complex NCOs. Table \ref{tab:streamed_data_options} outlines the streamed data options.

\begin{table}[h!]
    \centering
    \begin{tabular}{lll}
        \textbf{Interface} & \textbf{Complex channel bandwidth} & \textbf{Number of Channels} \\
        \toprule
        Standard 1G Ethernet & 596\,Hz / 1.192\,kHz / 2.384\,kHz & 4,096 \\
        100G\,QSFP28 & 2.44\,MHz (native channel output) & up to 512 \\
        100G\,QSFP28 & 625\,MHz (NCO output) & up to 4 \\
        \bottomrule
    \end{tabular}
    \vspace{1em} 
    \caption{Continuously streamed data output options.}
    \label{tab:streamed_data_options}
\end{table}

Discrete captures from the NCO or native 2.44\,MSPS channel data are possible through the 1\,Gbps Ethernet interface. This allows detailed characterization of systems without 100\,Gbps networking infrastructure. One example of using this feature is shown in Figure \ref{fig:kid-dashboard}, in which discrete captures at 2.44\,MSPS are used as part of a detector dashboard to identify the generation/recombination noise roll-off.

\subsection{Active feedback}
\label{sec:feedback}

Coupling the demodulation and synthesis PFBs enables output tones (frequency, phase, and amplitude) to be varied dynamically based on the incoming data.
The feedback core can implement a variety of feedback techniques with a basic structure of an integrating controller (Figure \ref{fig:feedback-diagram}) in which an accumulator sets an error function by which the complex output is adjusted, and complex offsets are used to determine the target set-point.
A similar signal processing technique was originally applied to the linearization of Superconducting Quantum Interference Devices (SQUIDs), a kind of superconducting device used in multiplexed arrays of Transition Edge Sensors.\cite{Smecher2022-dan,Smecher-litebird-firmware}

\begin{figure}
    \centering
    \includegraphics[width=0.5\linewidth]{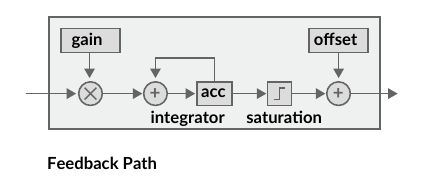}
    \caption{The basic block diagram of the feedback loop. Outputs are modified according to an integrating feedback loop configured with a (complex) gain and dynamically updated (complex) offset. The output waveform's amplitude, phase, and frequency are then modulated according to various linear combinations of input I and Q data.}
    \label{fig:feedback-diagram}
\end{figure}

The signal path structure is designed to be modified for a variety of feedback schemes, and t0.technology collaborates with academic researchers to support hardware implementations of novel control techniques, and to make the outcomes available to researchers. An active example is the case of Active Resonator Feedback (ARC). ARC is being pioneered by McGill researchers to address limitations with ``tone tracking'' feedback associated with inter-modulation distortion products. It was first described in Rouble et al. 2024 (these proceedings)\cite{rouble2024} to linearize detector response; extend dynamic range; and allow operation beyond traditional bifurcation powers. t0.technology is working with researchers to support real-time firmware implementations of ARC on CRS boards with the \texttt{rfmux} firmware, and to characterize the behavior of the detectors under that feedback.

\section{Application programming interface and data acquisition}

The primary application programming interface (API) used to interact with the CRS is an open-source Python library, which is used to issue commands and collect discrete data asynchronously across many CRS boards. That library is also embedded on the CRS boards and accessible through JupyterLab, intended as a simple way to operate individual CRS boards without a separate control computer or software dependencies.

The \texttt{rfmux} API represents CRS boards as Python classes with a collection of methods corresponding to the basic signal processing commands, such as setting NCO and channel parameters; retrieving samples from the different sampling domains; adjusting the the RF-DAC and RF-ADC digital gain, attenuation, and Nyquist region; selecting streaming rates; and parameterizing feedback paths. These methods can be combined to build custom routines, such as network analysis, parametric sweeps and tuning algorithms, or multi-probe investigations into superconducting device behavior. This provides a basic foundation of tools and routines which can be forked and extended for specialized applications. Extensions may include custom routines for application-specific hardware or instruments, data visualization, or combining analysis steps directly within algorithmic control software.

Distributed with \texttt{rfmux} is a simple C++ data acquisition (DAQ) system called the \texttt{parser}. The parser captures multicast UDP packets from the CRS, validates data integrity, and saves the data into a \texttt{dirfile}\footnote{\url{getdata.sourceforge.net/dirfile.html}} format that can be read by KST2\footnote{\url{kst-plot.kde.org/index.html}} or imported directly into Python. Most deployment instruments have existing DAQ pipelines, and so the \texttt{parser} also serves as an open-source ``how-to'' for integrating the UDP packet structure with 3rd party DAQ.

\subsection{Compatibility with third party software tools}

The CRS API can be embedded into other control software libraries. \texttt{hidfmux} is a general purpose KID readout software package developed by McGill University researchers that implements deployment-scale algorithms for resonator characterization, network analyses, and bias optimization strategies.\cite{Rouble2022} While \texttt{hidfmux} was originally developed for the RF-ICE readout system, the \texttt{rfmux} API has now been integrated such that the CRS can also be used. \texttt{hidfmux} is the KIDs control software used for the South Pole Telescope (SPT-)SLIM experiment,\cite{Karkare2022} and proposed SPT-3G+ experiment. Integration of \texttt{rfmux} API supports the proposed SPT-3G+ experiment, which has changed its readout baseline use CRS hardware for readout and control.\cite{Anderson2022}

In preparation for SPT-3G+, the open-source data acquisition software \texttt{ledgerman} will also be modified to include CRS compatibility. \texttt{ledgerman} was originally developed for the South Pole Telescope (3G) and subsequently open-sourced.\footnote{\url{https://cmb-s4.github.io/spt3g_software/dataacquisition.html}}

\section{Characterization and operation of KIDs}

The CRS operates with a different paradigm than traditional homodyne vector network analyzers (VNAs) or arbitrary waveform generators. The large number of independent sinusoidal tones available in each readout module can be simultaneously synthesized, demodulated, and adjusted in phase, amplitude, and frequency at low latency. 
They remain phase-coherent with each other, with tones from other I/O modules, from other CRS boards, and over time.

This can be exploited to accelerate parallelizable operations such as network analyses (Section \ref{sec:netanal}) or parametric sweeps of devices across a full array (Figure \ref{sec:characterization}). It can also be used to synchronously sample a wide instantaneous bandwidth in order to study superconducting system dynamics and perturbations in novel ways (Section \ref{sec:bifurcation}).

The results in this section are intended to illustrate how the CRS and \texttt{rfmux} firmware can be used. All figures are generated with \texttt{hidfmux} software, operating a CRS, with prototype detectors developed for the SPT-3G+ instrument provided by the South Pole Telescope collaboration.

\subsection{Contrast with homodyne VNA operations}
\label{sec:netanal}

The CRS signal processing is conceptually similar to treating each multiplexed channel as an independent parallel VNA. A network analysis to locate resonances across a wide bandwidth will synchronously sample all of those channels. In the default configuration, 1,024 tones are spaced across \SI{625}{\mega\hertz} of bandwidth and stepped by the target resolution. Additional \SI{625}{\mega\hertz} segments are measured in the same manner after incrementing the NCO. Coverage of \SI{1}{\giga\hertz} of bandwidth at $\sim$\SI{8}{\kilo\hertz} resolution uses 120,000 independent frequency points, corresponding to 118 synchronous measurements of the frequency comb and two NCO set points. This operation completes in under a minute using the \texttt{hidfmux} measurement algorithm (Figure \ref{fig:netanal}), in parallel for each I/O module and asynchronously across multiple CRS boards.\footnote{
Further optimization of these execution times is possible, as the fundamental limit for  parallel system is more than an order of magnitude faster.}
\begin{figure}
    \centering
    \includegraphics[width=0.75\linewidth]{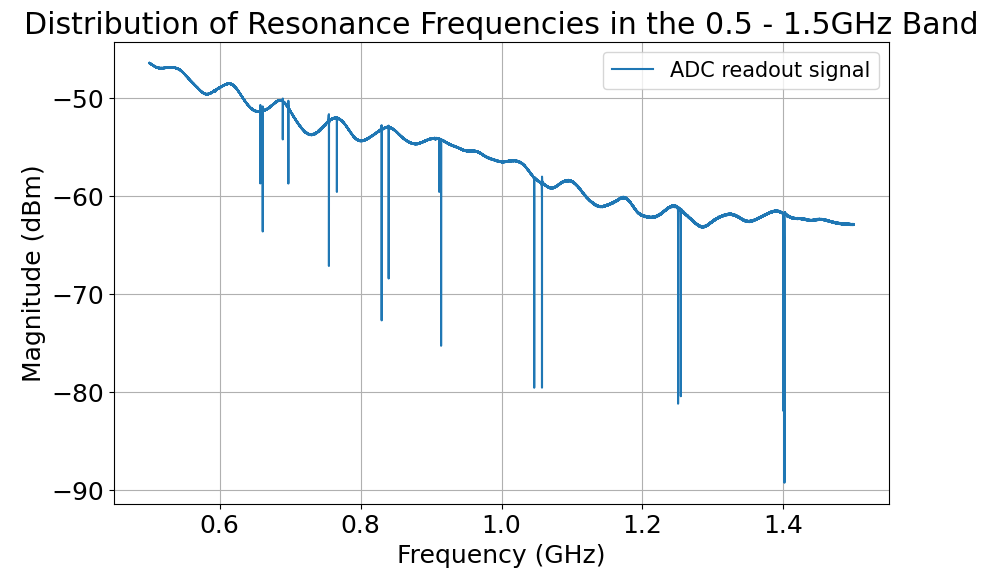}
    \vspace{1em} 
    \caption{A network analysis over \SI{1}{\giga\hertz} of bandwidth with $\sim$8 kHz resolution and 120,000 frequency points and 1.2 million samples. Each point is an average of 10 samples. The total execution time is under a minute. This sweep uses a testing wafer with sixteen KID resonators.}
    \label{fig:netanal}
\end{figure}

\subsection{KID characterization}
\label{sec:characterization}

A similar technique to the parallel network analysis is used to characterize individual KID resonances once their resonant frequencies have been coarsely located. Each resonance is allocated to a channel and simultaneous parametric sweeps across all resonances identify optimal bias frequencies and amplitudes. These data are combined with coincident 2.44\,MSPS captures at the selected bias point to characterize the noise properties of the system within the readout bandwidth.
The dashboard shown in Figure \ref{fig:kid-dashboard} is automatically generated with a \texttt{hidfmux} routine that follows this methodology, with an execution time dominated by the parallel acquisition of 60 seconds of data.

\begin{figure}
    \centering
    \includegraphics[width=1\linewidth]{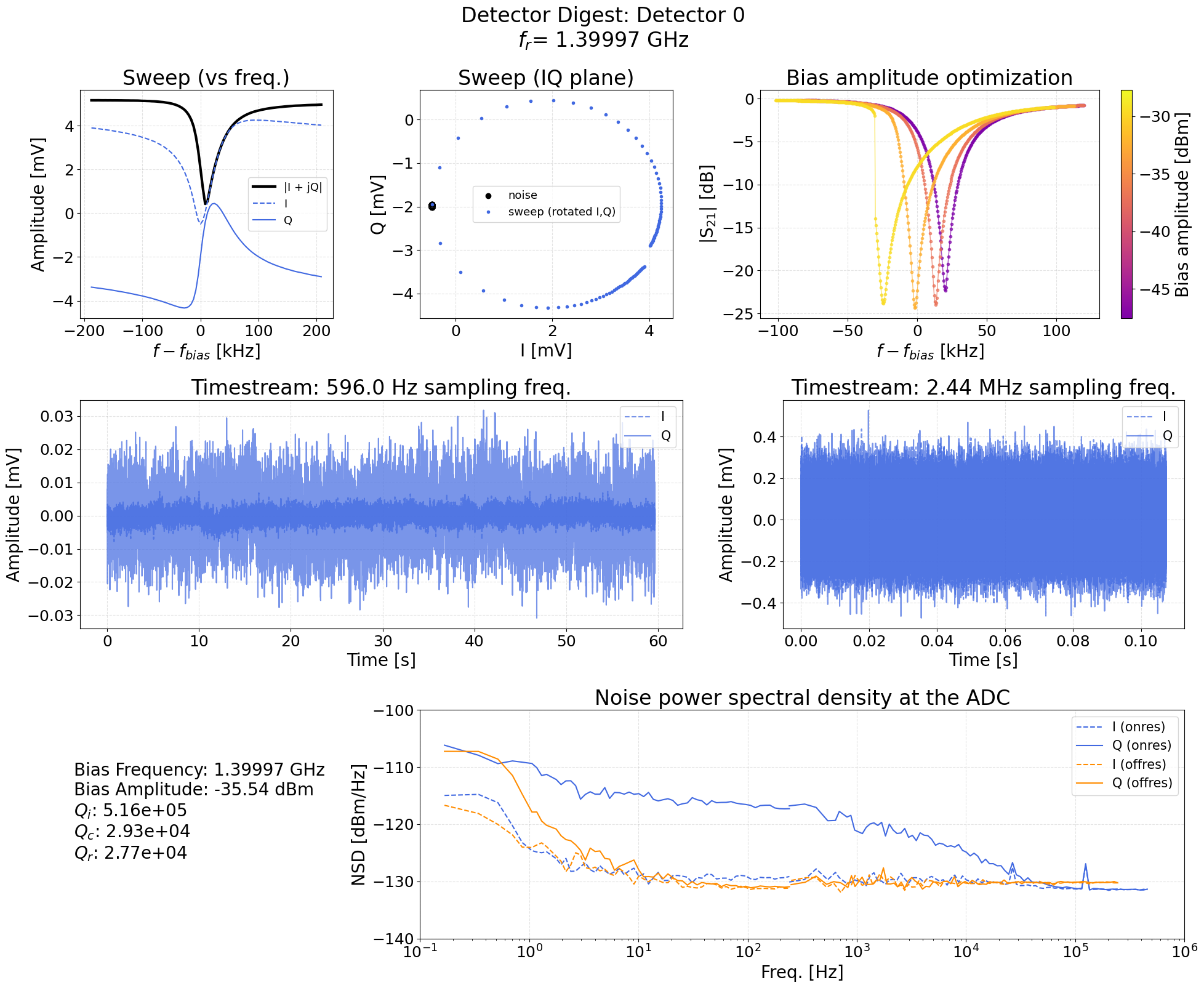}
    \vspace{1em} 
    \caption{Example KID characterization dashboard. This is an output of an initialization routine within \texttt{hidfmux} used to quickly evaluate and tune a KID array. It typically follows the coarse network analysis shown in Figure \ref{fig:netanal}, with the approximate resonance locations as an input. The algorithm determines the appropriate bias frequency and amplitude, and provides a noise snapshot using both the slow and fast sampling bandwidths.}
    \label{fig:kid-dashboard}
\end{figure}

\subsection{Multi-probe view of KID dynamics}
\label{sec:bifurcation}

Typically, KIDs are operated with a single tone per detector, such that the output of that single channel is used to infer the behavior of the superconducting device and resonance as a whole. However, those properties can be better understood with a coincident view across the full resonator bandwidth.

This is well illustrated in the case of bifurcation, a behavior in which the resonance distorts in the presence of a sufficiently large tone.\cite{deVisser2010} A VNA measurement of this behavior is limited to a single measurement at each frequency location (that of the tone) while monotonically sweeping that tone in frequency space. Seen this way, the resonance appears as in the top panel of Figure \ref{fig:bifurcation}, and the overall dynamics of the resonator must be inferred.
By incorporating a comb of probe channels at much lower amplitude (-40\,dBc), the full temporal evolution of the resonator across its full bandwidth is recorded, shown in the bottom panels of Figure \ref{fig:bifurcation}.

\begin{figure}[hbtp!]
    \centering
    \includegraphics[width=0.85\linewidth]{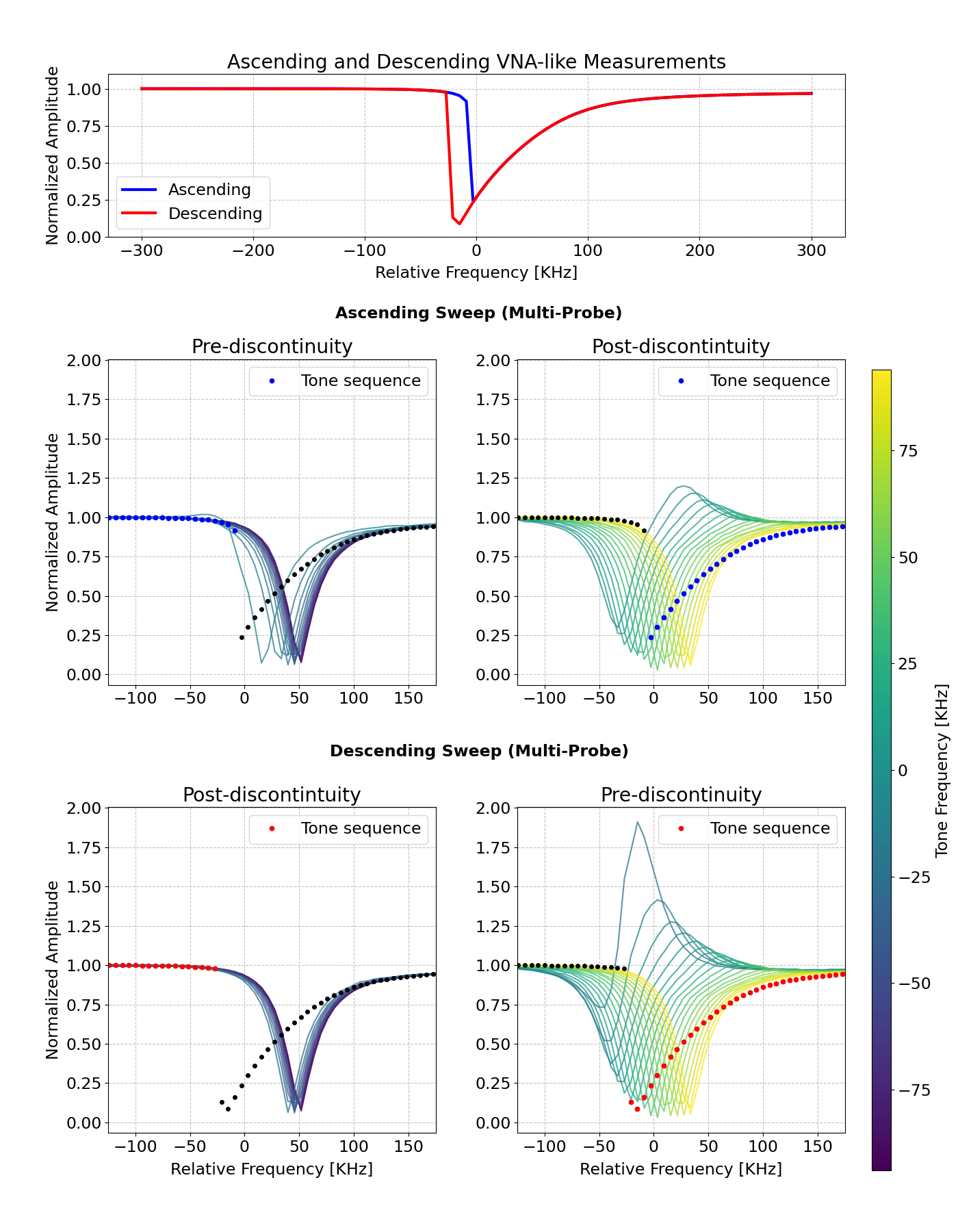}
    \caption{A comparison between a VNA-like measurement and multi-probe view of the dynamics of a Kinetic Inductance Detector under a large amplitude tone. The top panel is what can be seen with a VNA, while the bottom panels illustrate the full temporal evolution of the system over a wide instantaneous bandwidth, as seen in a multi-probe measurement. This measurement was conducted using a CRS board and the open-source \texttt{rfmux} API, executing a measurement algorithm in \texttt{hidfmux}.}
    \label{fig:bifurcation}
\end{figure}

The multi-probe measurement technique for characterizing the dynamics of KIDs was first described by a McGill research group\cite{rouble2024} while investigating detectors developed for South Pole Telescope. These measurements are now natively supported on CRS hardware and this figure was made using a CRS board and the open-source \texttt{rfmux} API, executing a measurement algorithm in \texttt{hidfmux}.

These measurements of KID dynamics are an example implementation of a general methodology that can be used to reconstruct the temporal evolution of superconducting systems under perturbations.

\section{Conclusion}

In this work, we introduce a new control system for superconducting devices. Focusing specifically on Kinetic Inductance Detector arrays, we provide the major technological parameters.
We describe the \texttt{rfmux} signal processing firmware, and illustrate some advantages over VNA-like systems for detector characterization and operation.

One of the objectives in building the CRS was to ensure availability of control electronics for superconducting detectors that are collaboration-agnostic, largely independent from individual academic project funding cycles, and with long-term dedicated hardware and engineering support.
The CRS is being introduced as a commercially available system under this paradigm. At the same time, it is intended to facilitate the kind of research that is best done without walled gardens. Much of the CRS development environment is open-source, including the \texttt{rfmux} API and control software.

This design philosophy informs the software interface (to be easily forked, adapted, and and integrated); the firmware signal path (accommodating both survey-style and particle-detection-style implementations); the hardware (wide bandwidth and an abundance of expandable peripherals without extraneous electromagnetic interference sources present on commercial evaluation platforms); and the deployment-scale ready systems integration (with custom subracks, high-speed intra-board connectivity, and phase synchronization).

This approach is also intended to facilitate open collaborations between t0.technology and researchers working on new ideas. This may include discussions regarding new measurements possible with the system; alternative ways to use the system with devices we haven't considered; collaboration for extending/validating support for open-source toolchains such as CASPER\cite{casper} or QICK\cite{Stefanazzi2022}; or participation in a collaborative research project with aligned research objectives.

\section*{CONTRIBUTIONS}

All measurements in this paper were taken by t0.technology affiliated authors.
The CRS hardware and electronics, \texttt{rfmux} firmware, and API software are all developed by the authors identified with t0.technology affiliations either directly through t0.technology or through internship programs with t0.technology.

Rouble et al.'s analysis of KIDs bifurcation behavior with the RF-ICE system\cite{rouble2024} motivated the application of multi-probe measurements highlighted in Section \ref{sec:bifurcation}; 
Rouble is the lead author of the \texttt{hidfmux} KID control software that produced Figures \ref{fig:netanal}, \ref{fig:kid-dashboard}, and \ref{fig:bifurcation}. \texttt{hidfmux} is a general purpose KID readout software package developed for the South Pole Telescope, and was not developed by t0.technology.

\acknowledgments 

All measurements of KIDs use a prototype SPT-3G+ detector wafer designed and fabricated by Kyra Fichman and Karia Dibert at the University of Chicago, which is on loan from the South Pole Telescope collaboration.
\\\\
This work was supported by Mitacs through the Mitacs Accelerate program.

\bibliography{main} 

\begin{thebibliography}{10}

\bibitem{Anderson2022}
Anderson, A.~J., Barry, P., Bender, A.~N., Benson, B., Bleem, L., Carlstrom,
  J.~E., Cecil, T.~W., Chang, C.~L., Crawford, T.~M., Dibert, K.~R., Dobbs,
  M.~A., Fichman, K., Halverson, N.~W., Holzapfel, W.~L., Hryciuk, A., Karkare,
  K.~S., Li, J., Lisovenko, M., Marrone, D., McMahon, J., Montgomery, J.,
  Natoli, T., Pan, Z., Raghunathan, S., Reichardt, C.~L., Rouble, M.,
  Shirokoff, E., Smecher, G., Stark, A.~A., Vieira, J.~D., and Young, M.~R.,
  ``Spt-3g+: mapping the high-frequency cosmic microwave background using
  kinetic inductance detectors,'' in [{\em Millimeter, Submillimeter, and
  Far-Infrared Detectors and Instrumentation for Astronomy
  XI}{\nolinebreak\hspace{0.1em}]},  Zmuidzinas, J. and Gao, J.-R., eds., SPIE
  (Aug. 2022).

\bibitem{Day2003}
Day, P.~K., LeDuc, H.~G., Mazin, B.~A., Vayonakis, A., and Zmuidzinas, J., ``A
  broadband superconducting detector suitable for use in large arrays,'' {\em
  Nature}~{\bf 425},  817–821 (Oct. 2003).

\bibitem{Park2022}
Park, K.~H., Yap, Y.~S., Tan, Y.~P., Hufnagel, C., Nguyen, L.~H., Lau, K.~H.,
  Bore, P., Efthymiou, S., Carrazza, S., Budoyo, R.~P., and Dumke, R.,
  ``Icarus-q: Integrated control and readout unit for scalable quantum
  processors,'' {\em Review of Scientific Instruments}~{\bf 93} (Oct. 2022).

\bibitem{Fruitwala2020}
Fruitwala, N., Strader, P., Cancelo, G., Zmuda, T., Treptow, K., Wilcer, N.,
  Stoughton, C., Walter, A.~B., Zobrist, N., Collura, G., Lipartito, I.,
  Bailey, J.~I., and Mazin, B.~A., ``Second generation readout for large format
  photon counting microwave kinetic inductance detectors,'' {\em Review of
  Scientific Instruments}~{\bf 91} (Dec. 2020).

\bibitem{Smecher2022-dan}
Smecher, G.~M., de~Haan, T., Dobbs, M., and Montgomery, J., ``Digital active
  nulling for frequency-multiplexed bolometer readout: performance and
  latency,'' in [{\em Millimeter, Submillimeter, and {Far-Infrared} Detectors
  and Instrumentation for Astronomy {XI}}{\nolinebreak\hspace{0.1em}]},
  Zmuidzinas, J. and Gao, J.-R., eds., SPIE (Aug. 2022).

\bibitem{Smecher-litebird-firmware}
Smecher, G.~M., Cliche, J.-F., Dobbs, M., and Montgomery, J., ``Development of
  trl5 firmware for tuning, biasing, and readout of kilopixel tes bolometer
  arrays,'' in [{\em Millimeter, Submillimeter, and Far-Infrared Detectors and
  Instrumentation for Astronomy XI}{\nolinebreak\hspace{0.1em}]},  Zmuidzinas,
  J. and Gao, J.-R., eds., SPIE (Aug. 2022).

\bibitem{rouble2024}
Rouble, e.~a., ``Large-scale frequency-multiplexed readout of superconducting
  resonators with rf-ice,'' in [{\em Proceedings of the SPIE Astronomical
  Telescopes and Instrumentation}{\nolinebreak\hspace{0.1em}]},  (2024).
\newblock In this proceedings.

\bibitem{Rouble2022}
Rouble, M., Smecher, G.~M., Anderson, A.~J., Barry, P.~S., Dibert, K., Dobbs,
  M., Karkare, K.~S., and Montgomery, J., ``Rf-ice: large-scale gigahertz
  readout of frequency-multiplexed microwave kinetic inductance detectors,'' in
  [{\em Millimeter, Submillimeter, and Far-Infrared Detectors and
  Instrumentation for Astronomy XI}{\nolinebreak\hspace{0.1em}]},  Zmuidzinas,
  J. and Gao, J.-R., eds., SPIE (Aug. 2022).

\bibitem{Karkare2022}
Karkare, K.~S., Anderson, A.~J., Barry, P.~S., Benson, B.~A., Carlstrom, J.~E.,
  Cecil, T., Chang, C.~L., Dobbs, M.~A., Hollister, M., Keating, G.~K.,
  Marrone, D.~P., McMahon, J., Montgomery, J., Pan, Z., Robson, G., Rouble, M.,
  Shirokoff, E., and Smecher, G., ``Spt-slim: A line intensity mapping
  pathfinder for the south pole telescope,'' {\em Journal of Low Temperature
  Physics}~{\bf 209},  758–765 (Mar. 2022).

\bibitem{deVisser2010}
de~Visser, P.~J., Withington, S., and Goldie, D.~J., ``Readout-power heating
  and hysteretic switching between thermal quasiparticle states in kinetic
  inductance detectors,'' {\em Journal of Applied Physics}~{\bf 108} (Dec.
  2010).

\bibitem{casper}
Hickish, J., Abdurashidova, Z., Ali, Z., Buch, K.~D., Chaudhari, S.~C., Chen,
  H., Dexter, M., Domagalski, R.~S., Ford, J., Foster, G., George, D.,
  Greenberg, J., Greenhill, L., Isaacson, A., Jiang, H., Jones, G., Kapp, F.,
  Kriel, H., Lacasse, R., Lutomirski, A., MacMahon, D., Manley, J., Martens,
  A., McCullough, R., Muley, M.~V., New, W., Parsons, A., Price, D.~C.,
  Primiani, R.~A., Ray, J., Siemion, A., Van~Tonder, V., Vertatschitsch, L.,
  Wagner, M., Weintroub, J., and Werthimer, D., ``A decade of developing
  radio-astronomy instrumentation using casper open-source technology,''
  (2016).

\bibitem{Stefanazzi2022}
Stefanazzi, L., Treptow, K., Wilcer, N., Stoughton, C., Bradford, C., Uemura,
  S., Zorzetti, S., Montella, S., Cancelo, G., Sussman, S., Houck, A., Saxena,
  S., Arnaldi, H., Agrawal, A., Zhang, H., Ding, C., and Schuster, D.~I., ``The
  qick (quantum instrumentation control kit): Readout and control for qubits
  and detectors,'' {\em Review of Scientific Instruments}~{\bf 93} (Apr. 2022).

\end{thebibliography}
\bibliographystyle{spiebib} 

\end{document}